\documentclass[graybox]{svmult}

\usepackage{mathptmx}       
\usepackage{helvet}         
\usepackage{courier}        
\usepackage{type1cm}        
\usepackage{amsmath}
\usepackage{mathtools}                            
\usepackage{makeidx}         
\usepackage{graphicx}        
\usepackage{multicol}        
\usepackage[bottom]{footmisc}

\usepackage{subcaption}
\makeindex             

\begin{document}

\title*{Resistance of communities against disinformation}
\author{Amirarsalan Rajabi, Seyyedmilad Talebzadehhosseini, and Ivan Garibay}
\institute{Amirarsalan Rajabi \at Complex Adaptive Systems Lab (CASL), University of Central Florida, Orlando, 32816\\ \email{amirarsalan@knights.ucf.edu}
\and Seyyedmilad Talebzadehhosseini \at Complex Adaptive Systems Lab (CASL), University of Central Florida, Orlando, 32816\\ \email{talebzadeh@knights.ucf.edu}
\and Ivan Garibay \at Complex Adaptive Systems Lab (CASL), University of Central Florida, Orlando, 32816\\ \email{igaribay@ucf.edu}}
%
%
\maketitle

\abstract{The spread of disinformation is considered a big threat to societies and has recently received unprecedented attention. In this paper we propose an agent-based model to simulate dissemination of a conspiracy in a population. The model is able to compare the resistance of different network structures against the activity of conspirators. Results show that \textit{connectedness} of network structure  and \textit{centrality} of conspirators are of crucial importance in preventing conspiracies from becoming widespread.}

\section{Introduction}
\label{sec:1}
We  define  conspiratorial  thinking  as  a  belief  held  by  an individual  or  a  group  of  individuals,  while  there  is  enough evidence  and  information  to  undermine  or  totally  refute  the belief. If some conspiracy theories get popular support, it may cause serious concerns.This is particularly true of conspiracies over  scientific  and  medical  issues  where  conspiracy  theories can result in rejection of the scientific method \cite{lewandowsky2013role}. 

Several  underlying  reasons  have  been  proposed  to  explain the  existence  of  conspiracy  theories.  According  to  Barkun, conspiratorial  thinking  exhibits  three  characteristics:  Firstly, nothing happens by accident. Secondly, things are not as they seems  on  surface.  And  thirdly,  things  are  highly  connected \cite{barkun2013culture}.  All  three  characteristics  mentioned  by  Barkun,  refer  to a  special  cognitive  function  of the  conspiracy  theorist.  Indeed, a  great  deal  of  literature  on  conspiracy  theory  associates conspiratorial  thinking  with  a  special  and  different  heuristic of the conspiracy theorist. On the other hand, Sunstein and Vermeule claim that many of those who hold conspiracy theories do so not as a result of a mental illness of any kind, or of simple irrationality, but as a  result  of  crippled  epistemology  (knowing  very  few  things, which are indeed wrong) \cite{sunstein2009conspiracy}. For most of what they believe and  know,  human  beings  lack  direct  information;  they  must rely on what others say and think. Hardin argues many people suffer from crippled epistemology, meaning they only get their information from a few incorrect sources \cite{breton2002political}. Crippled epistemology usually takes place in echo chambers. Echo   chambers   are   communities   in   which   individuals merely communicate with each other and rarely seek information from entities outside the community. The advent of social media  platforms  has  resulted  in  the  rise  of  echo-chambers \cite{bakshy2015exposure}.  Echo  chambers  play  an  important  role  in  political  and social  polarization  \cite{barbera2015tweeting}. The negative effects of polarization in social networks is studied in \cite{garibay2019polarization}. Bauman  states  that individuals  who  are  embedded  in  isolated  groups  or  small, self-enclosed  networks  who  are  thus exposed  only  to  skewed  information,  will  more  often  hold  conspiracy  theories  that  are justified,  relative  to  their  limited  informational  environment \cite{bauman2013liquid}.  The  study  of the dynamics by which the echo chambers form,  can  therefore  shed  light  on the  mechanisms  by  which  conspiratorial  thinking  forms  and thrives  in  a  community.

Traditional  studies  of  conspiratorial  thinking  assume  that conspiratorial thinking extinguishes in large network structure \cite{grimes2016viability},  and  that  conspiratorial  ideation  is  because  of  flawed reasoning  and  biased  heuristics  \cite{barkun2013culture}.  This  study  challenges both of these claims. Contrary to a large body of literature on conspiracy theory that  studies  the cognitive  function of isolated individuals [22],  this paper takes into account systemic belief dissemination as a result of interactions between individuals.

\section{Opinion Dynamics}
Opinion formation is a complex process which is formed by the interaction of multiple underlying elements. People tend to form their opinion on a wide variety of subjects through the process of learning. “Social learning” is a term referring to the process of learning through the communication of individuals with each other, their own experience and their observations of others’ experiences, media sources, propaganda and indoctrination from political leaders and the state \cite{acemoglu2011opinion}. In this paper, we refer to models of opinion dynamics as mathematical models that aim at capturing the dynamics of social learning, opinion spreading, collective decision making, and so on from a mathematical point of view.

Models of opinion dynamics can be divided into two categories : Bayesian models of opinion dynamics and non-Bayesian models of opinion dynamics. Bayesian models rely on Bayes rule \cite{bayes1991essay}. These kinds of models assume that an individual (agent) is Bayesian rational and update their belief optimally with respect to Bayes rule, given an underlying model of the world. One problem with Bayesian models of opinion dynamics is that these models make a lot of assumptions. One demanding assumption of these models is that an agent must have a reliable prior assumption about the world, an assumption that might be unrealistic in many cases. Additionally, it is assumed that they then go on and update their prior beliefs based on the new information that they get from others. Bayesian models also put too much structure on updating by ruling out “zero probability events” \cite{acemoglu2011opinion}. The aforementioned problems make Bayesian models unfavorable to be incorporated in our study of dissemination of conspiracy theories.    

Non-Bayesian approaches and models on the other hand try to avoid some of these problems. Non-Bayesian approaches are believed to be more effective in modeling belief manipulation and the spread of disinformation. The simplest of these models start by specifying rules of thumb \cite{acemoglu2011opinion}. Several different non-Bayesian models exist. Classical models of interacting particle systems which in inspired by statistical mechanics inspires many of these models (see for example see  for  example  \cite{clifford1973model}; \cite{galam2002minority}; \cite{latane1981psychology}; \cite{castellano2009nonlinear}; \cite{hegselmann2002opinion}). 

One noteworthy non-Bayesian model is DeGroot (1974) \cite{degroot1974reaching}. In this model, a set of interacting agents start by an initial beliefs about an underlying state held by each agent, and exchange information about their beliefs with their neighbors and update their beliefs at discrete time instances, with respect to a weight matrix that represents the social network structure of interactions. This model captures the “imitation” aspect of non-Bayesian models. While notably innovative, this model suffers from duplication of information \cite{acemoglu2010spread}, meaning that agents in this model might interact endlessly with their neighbors that hold unchanging opinions in each timestep.

\section{Model of Conspirators}
Our agent-based model is inspired by the variation of Acemoglu et. al. on DeGroot model \cite{acemoglu2010spread}. In our model, two types of agents exist. Majority of agents are \emph{susceptible} agents and a minority of agents exist that are called \emph{conspirators}. Conspirators deliberately disseminate false information to susceptibles.

Let's consider a conceptual underlying state of the world and call it $\Theta$. We assume that the true value of $\Theta$ is 1, and the discussion between agents in the model is on the true value of $\Theta$. $x_{i}^{s}(k)$ and $x_{i}^{c}(k)$ represent the opinion of susceptible agent $i$ at time $k$, and the opinion of conspirator agent $i$ at time $k$, respectively.
At first, each susceptible agent holds an initial belief about the underlying state. \textbf{\textit{N}} shows the total number of agents in the model. The initial belief of each susceptible agent is a randomly generated float number between zero and two, the initial belief of each conspirator agent is 0, and the average of initial beliefs of all susceptible agents is very close to one:

\begin{equation}
x_{i}^{s}(0) \in [0,2] \qquad x_{i}^{c}(0) = 0 \qquad \frac{1}{N}\sum_{i}x_{i}^{s}(0)\approx 1
\end{equation}

Hence although at first each susceptible agent has their own initial belief on the underlying state $\Theta$, there is a consensus on this underlying state between the susceptible agents. On the other hand, the initial belief of conspirator agents is 0 and remains 0 during the simulation, irrespective of their interactions. These could be thought of as individuals, entities, media or propaganda outlets that deliberately and constantly disseminate false information throughout the population. 

\begin{figure}[!b]
	\begin{subfigure}[b]{0.5\textwidth}
		\includegraphics[width=0.95\linewidth]{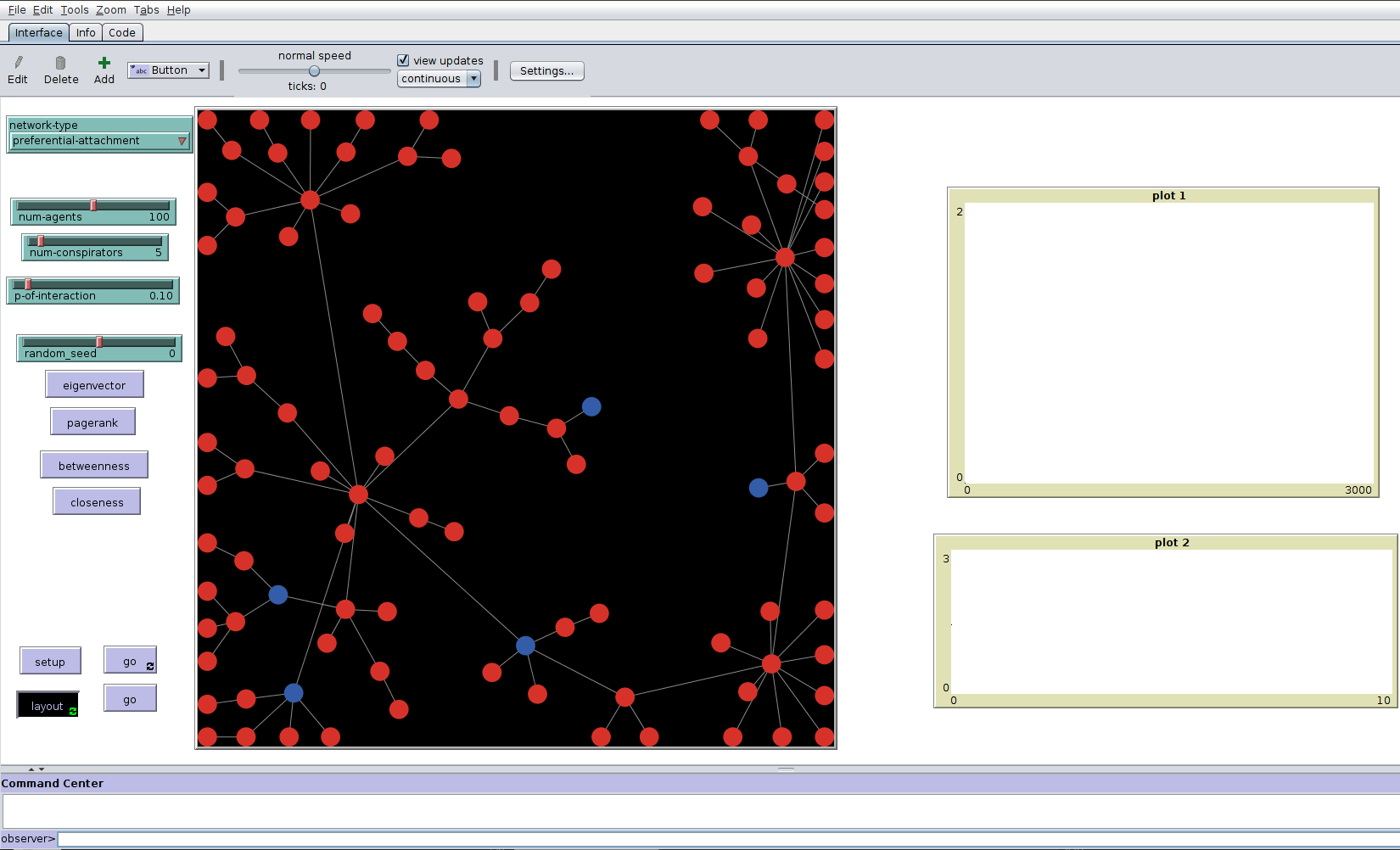}
		\caption{Scale Free (Barabasi-Albert)}
		\label{fig:gull}
	\end{subfigure}%
	\begin{subfigure}[b]{0.5\textwidth}
		\includegraphics[width=0.95\linewidth]{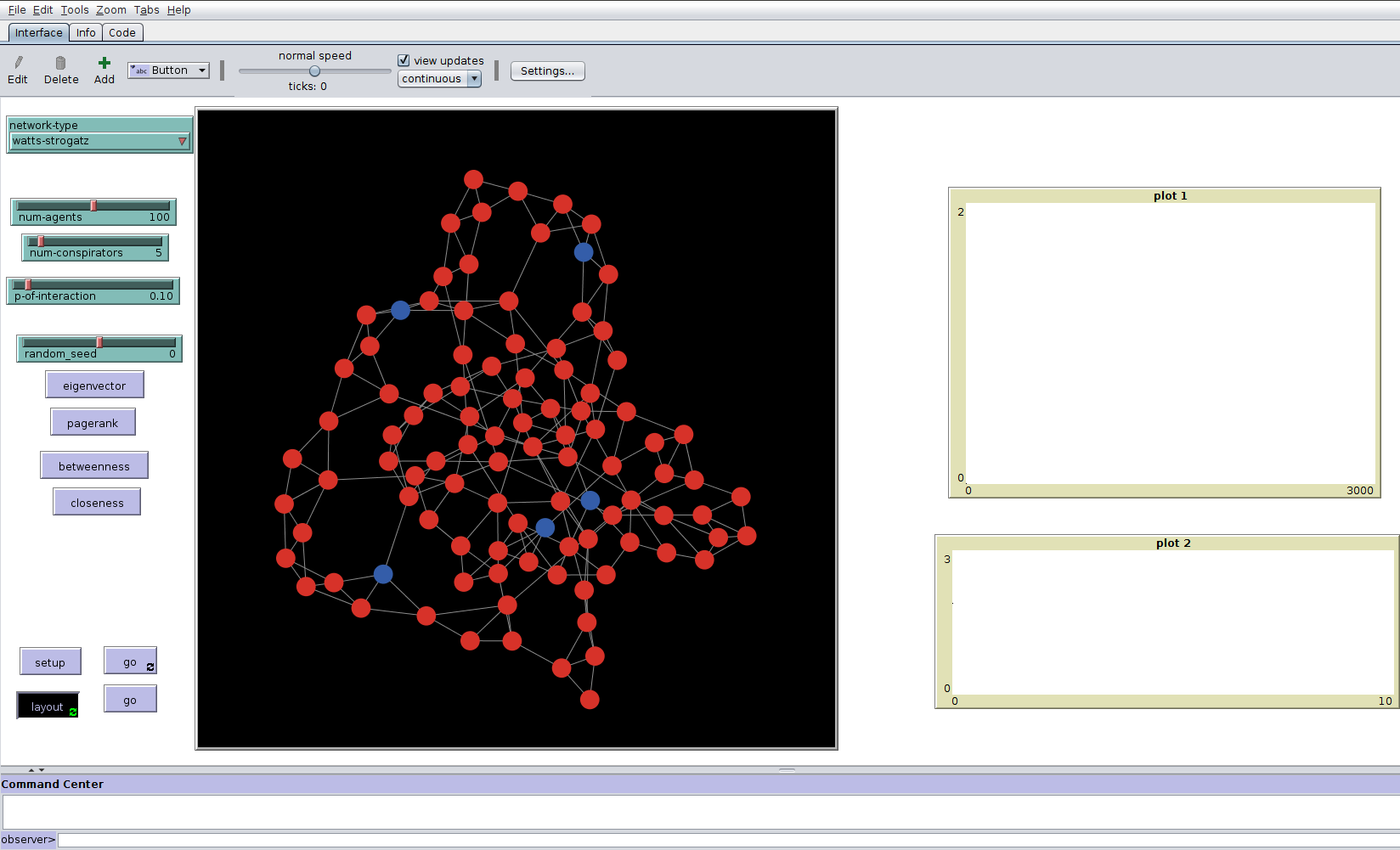}
		\caption{Small World (Watts-Strogatz)}
		\label{fig:gull2}
	\end{subfigure}%
	\caption{NetLogo environment. A) shows a network generated with Barabasi-Albert algorithm. B) shows a network generated by Watts-Strogat algorithm}
	\label{fig:environment}
\end{figure}

There are two main differences of between our proposed model and the work of \cite{acemoglu2010spread}: 1) Agents are not able to freely communicate directly with any other agents. Rather, each agent is only able to communicate with the agents that she is connected to by a link. This means that a network structure regulates the communication in the population. 2) In the variation of \cite{acemoglu2010spread} on DeGroot model, each agent meets and communicates with other agents at instances defined by a rate one Poisson process independent of other agents. In our model agents are connected to each other by undirected links, and together, they form a network. in each timestep of the simulation, an agent will choose another agent to which it is connected and may or may not communicate with them. In this model, the capacity of an agent for communication in each timestep is the number of her links. For example an agent with 5 links will communicate with 5 or less other agents in each timestep.  

Following the interaction of two agents i and j, there is a potential exchange of information between them with probability $\textrm{p}_{interaction}$. Agents update their beliefs according to one of the following possibilities:

\begin{equation}
\begin{aligned}
\nonumber
& \textrm{If \textit{i} and \textit{j} are both susceptibles:}\\
& \begin{cases}
x_{i}(k+1) = x_{j}(k+1) = \frac{1}{2}[x_{i}(k) + x_{j}(k)] & \text{with probability}\; \textrm{p}\\
x_{i}(k+1)=x_{i}(k)\; \& \; x_{j}(k+1)=x_{j}(k)  & \text{with probability}\; 1-\textrm{p}
\end{cases}\\ \\
& \textrm{If \textit{i} is susceptible and \textit{j} is conspirator:}\\
& \begin{cases}
x_{i}(k+1) = \frac{x_{i}(k) + 0}{2} & \text{with probability}\; \textrm{p}\\
x_{i}(k+1)=x_{i}(k)  & \text{with probability}\; 1-\textrm{p}\\
x_{j}(k+1)=x_{j}(k)=0 &  \text{with probability 1}
\end{cases}\\ \\
& \textrm{If \textit{i} and \textit{j} are both conspirators: no opinion sharing.}\\ \\
\end{aligned}
\end{equation}

The underlying network structure of the model plays an important role in determining the behavior of the model, and it is itself determined by the algorithm by which the model forms the network. It determines which agents are connected to each other, how easily the information would propagate throughout the network, and so forth.

Watts and Strogatz proposed a model to capture the small world, high clustering, and low average path properties of real complex networks \cite{watts1998collective}. On the other hand, Barabasi and Albert showed that power-law degree distribution is the property of many real world networks and proposed an algorithm that can capture this phenomenon \cite{barabasi2003scale}. Indeed, it is believed that majority of complex networks exhibit small world and scale free properties \cite{wang2003complex}. In order to capture all of these essential properties, the model generates the network of agents using both Watts-Strogatz and Barabasi-Albert algorithms. 

\section{Results}
\begin{figure}[t]
	\begin{subfigure}[b]{0.5\textwidth}
		\includegraphics[width=0.95\linewidth]{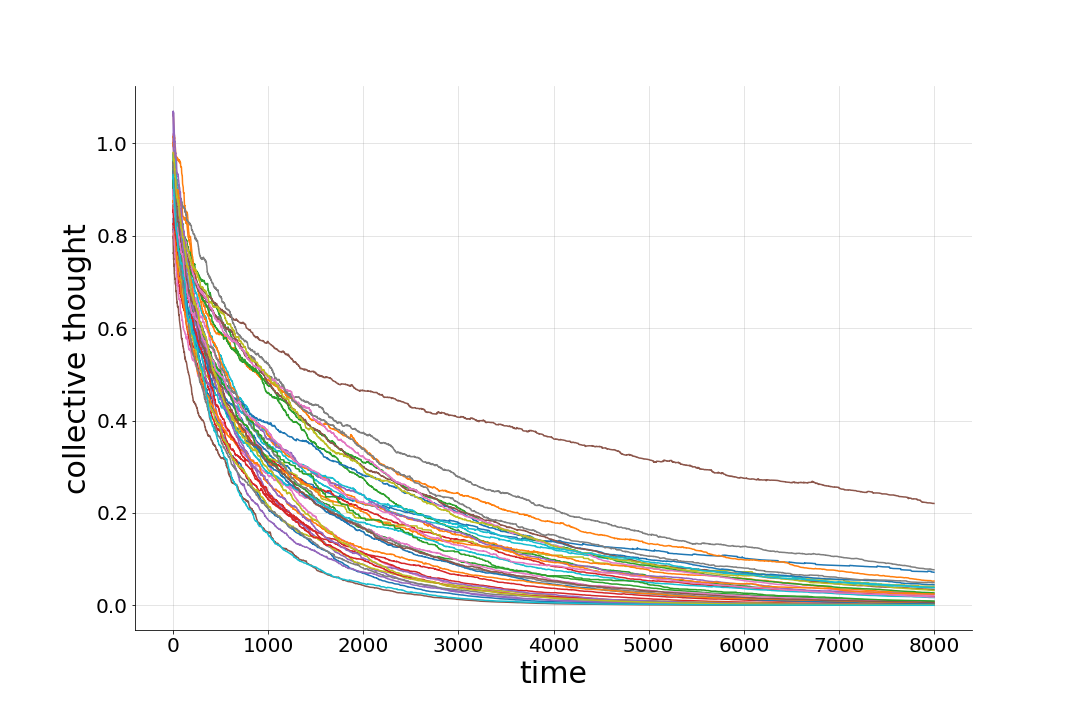}
		\caption{Scale Free (Barabasi-Albert)}
		\label{fig:gull}
	\end{subfigure}%
	\begin{subfigure}[b]{0.5\textwidth}
		\includegraphics[width=0.95\linewidth]{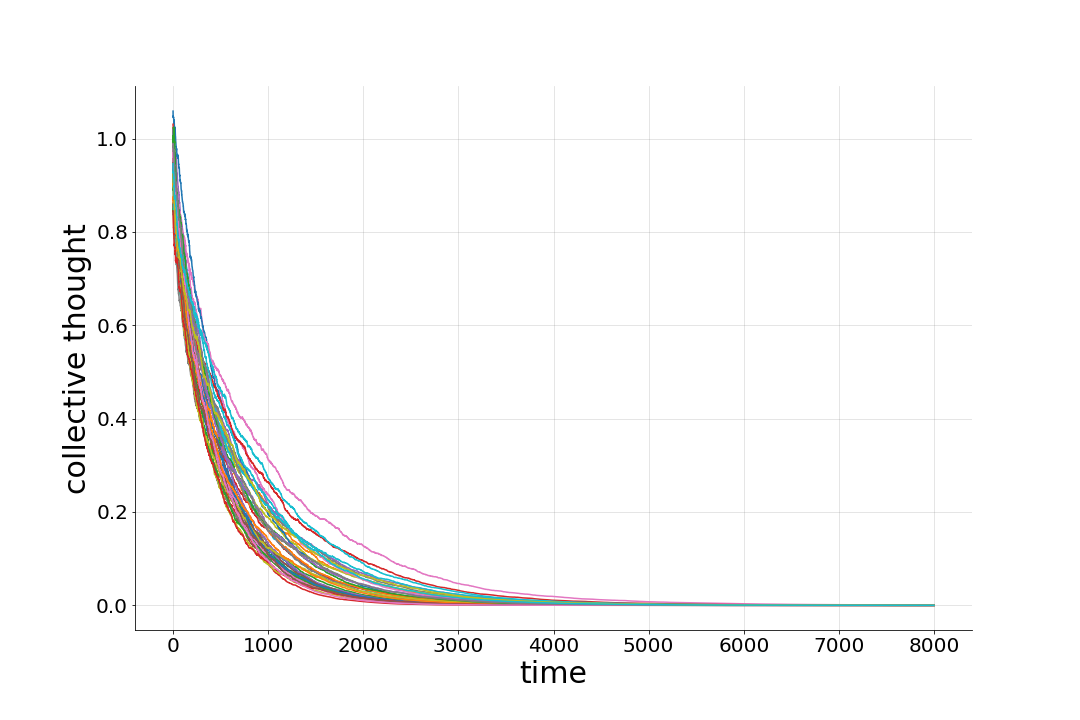}
		\caption{Small World (Watts-Strogatz)}
		\label{fig:gull2}
	\end{subfigure}%
	\caption{Figure show the required timesteps required for the network's collective-thought to converge. Small world network tends to be deceived more quickly than a scale free network.}
	\label{fig:collective}
\end{figure}

We developed our agent-based model in NetLogo \cite{sunstein2009conspiracy}. Figure~\ref{fig:environment} shows two instances of the model with watts-strogatz and Barabasi-Albert model.
For our purposes we defined a variable named collective-thought. This variable is simply the average of the belief of all susceptible agents and represents the collective belief of the population about the underlying state $\Theta$. Collective-thought starts with a value close to one, and always converges to zero(Figure ~\ref{fig:collective}). 

A desired population is one that resists the activity of conspirators. Numerous individuals and groups deliberately disseminate false information in a society. As we know the collective-thought of the model always converges to zero after some timesteps. We call this state the convergence. In order to quantify the resistance of a network against the activity of conspirators, we record the number of timesteps required for collective-thought of the network to converge. This number can be thought of as symbolically representing how long a society resist the effort of conspirators, before it is nearly deceived into believing that the underlying state $\Theta$ is zero.

We argue that contrary to conventional belief that conspiratorial beliefs are untenable with larger network structure (e.g. \cite{grimes2016viability}), a large network cannot ensure eradication of conspiratorial beliefs. Indeed, the important aspect of a network structure that can ensure resistance against conspiracies is the \emph{connectedness} of the network;that is, how easily an agent can send an information through the network to any other agent. A network with high connectedness means that information propagates more easily and eco-chambers are less likely to form.  We needed to come up with a benchmark that enables us to quantitatively compare different networks by our objective of connectedness. For this purpose, the model records the mean path length of the network in each instance of the simulation as follows \cite{fronczak2004average}:

\begin{equation}
\textrm{mean path length = average shortest path between all distincrt pairs of nodes in network}
\end{equation}

\begin{figure}[t]
	\begin{subfigure}[b]{0.5\textwidth}
		\includegraphics[width=0.95\linewidth]{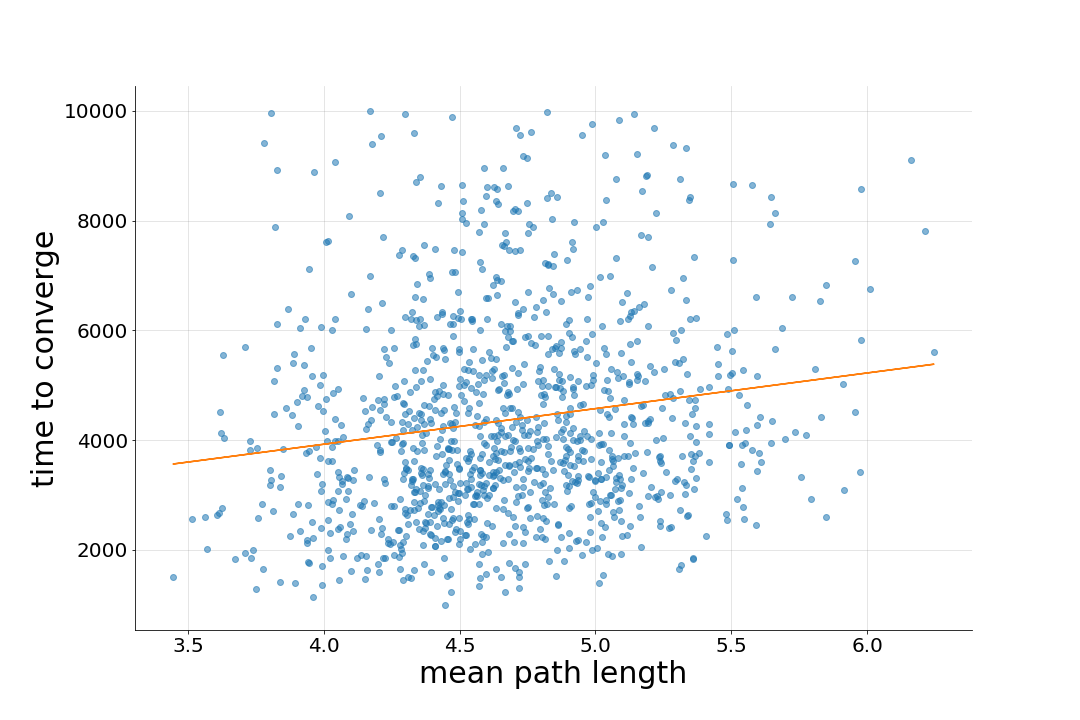}
		\caption{Scale Free (Barabasi-Albert)}
		\label{fig:gull}
	\end{subfigure}%
	\begin{subfigure}[b]{0.5\textwidth}
		\includegraphics[width=0.95\linewidth]{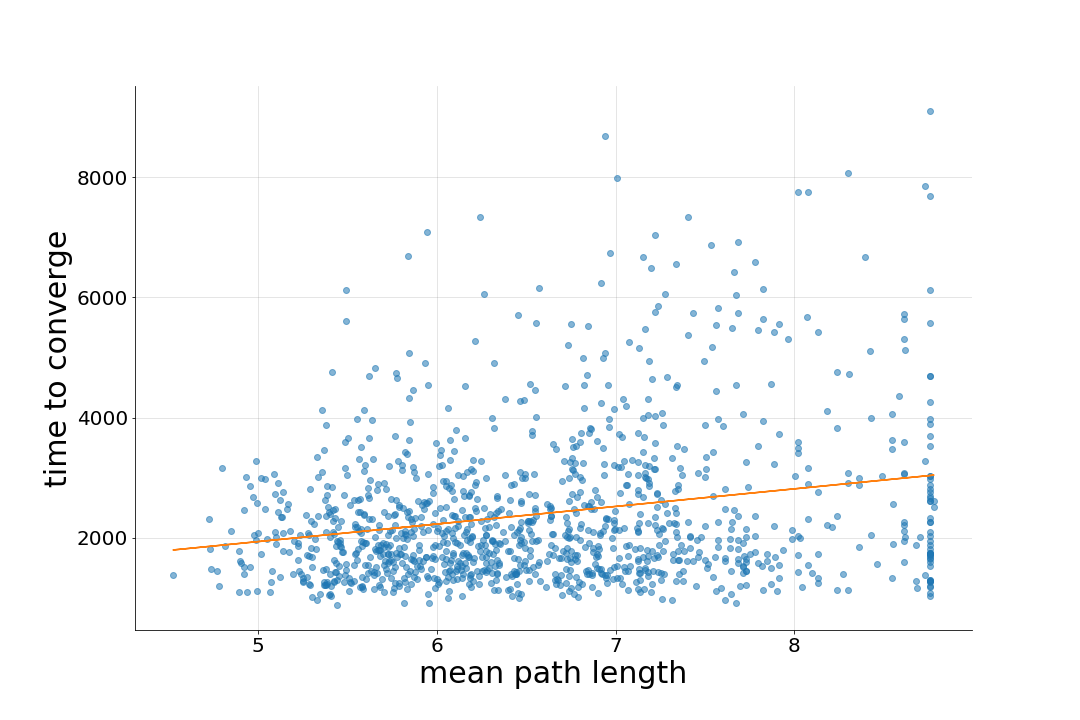}
		\caption{Small World (Watts-Strogatz)}
		\label{fig:gull2}
	\end{subfigure}%
	\caption{The relationship between mean path length and the required timesteps for a population's collective-thought to converge in A)scale free and B)small world networks}
	\label{fig:meanpath}
\end{figure}

The model was run 1000 times for each of the Watts-Strogatz and Barabasi-Albert networks. In each run, the required timsteps for the collective-thought to converge and the mean path length of the network were recorded. Figure~\ref{fig:meanpath} shows that there is a positive relationship between mean path length of the network, and required timesteps for the collective-thought to reach convergence. With Barabasi-Albert algorithm, the Pearson correlation coefficient between the mean path length and the required timesteps was 0.18, with a p-value less than 0.00001 which shows the correlation is statistically significant. The corresponding coefficient for Watts-Strogatz networks was 0.39 with a p-value less than 0.00001, which shows this correlation is also significant. These results prove our hypothesis that the connectedness of a network improves its resistance against conspirators.

\begin{figure}[!b]
	\begin{subfigure}[b]{0.5\textwidth}
		\includegraphics[width=0.95\linewidth]{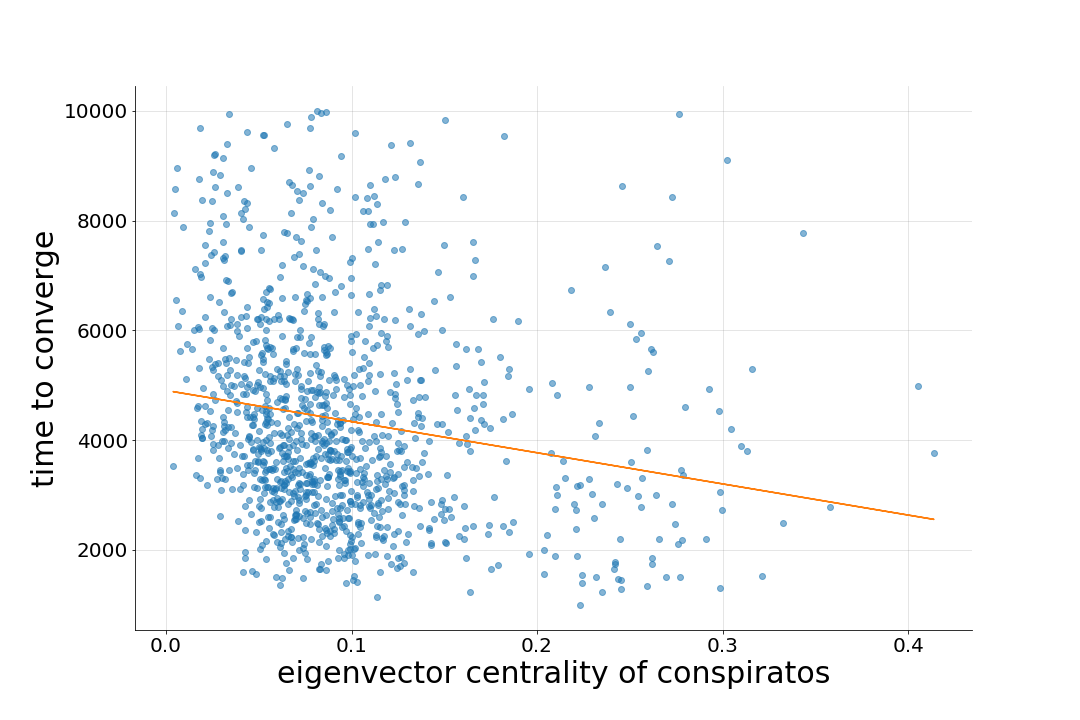}
		\caption{Scale Free (Barabasi-Albert)}
		\label{fig:gull}
	\end{subfigure}%
	\begin{subfigure}[b]{0.5\textwidth}
		\includegraphics[width=0.95\linewidth]{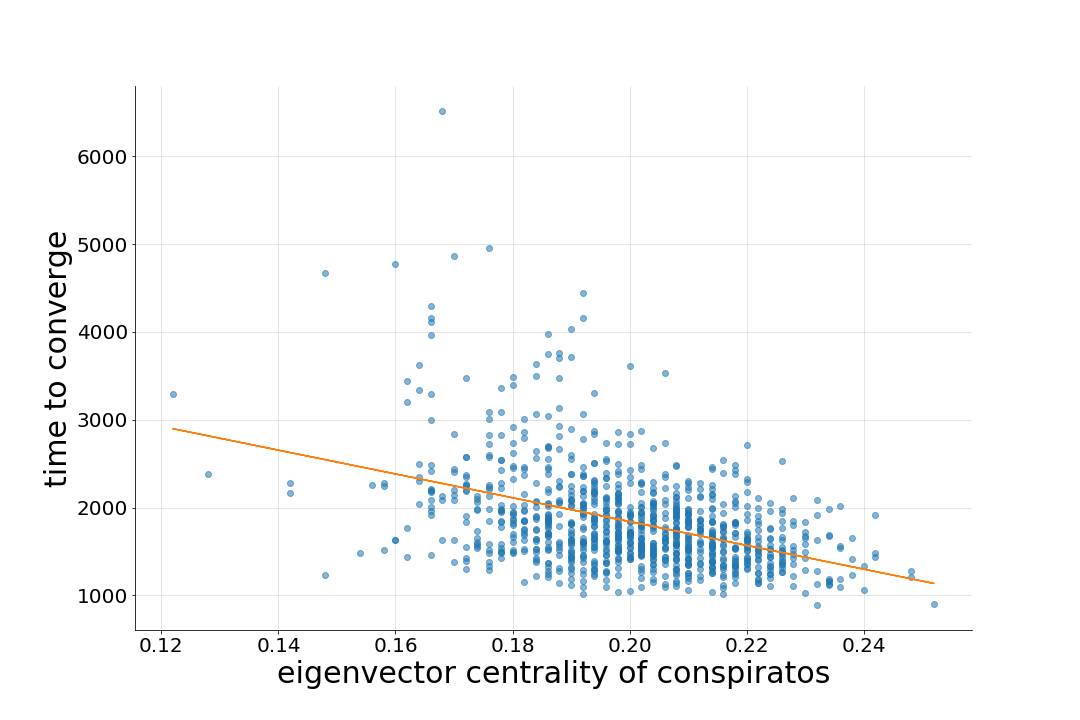}
		\caption{Small World (Watts-Strogatz)}
		\label{fig:gull2}
	\end{subfigure}%
	\caption{The relationship between sum of the eigenvector of all 4 conspirators and the required timesteps for a population's collective-thought to converge in A)scale free and B)small world networks}
	\label{fig:closeness}
\end{figure}

Besides the connectedness of a network, the structural position of conspirators in a network determines the reach of their disinformation and enhances their ability to disseminate their message through the whole population more effectively. To show that our model captures this phenomenon, we assigned to each agent in the network their eigenvector centrality. Eigenvector centrality is a measure of centrality in a network \cite{newman2008mathematics}. The more central a node is, the closer it is to all other nodes.

The model was run 1000 times for each of the Watts-Strogatz and Barabasi-Albert networks. For each run, the required timsteps for the collective belief to reach convergence and the sum of the eigenvector centrality of conspirator agents were recorded. Figure~\ref{fig:closeness} shows the negative relationship between required timesteps for collective-thought to reach convergence and the sum of the eigenvector centrality of conspirators. With Barabasi-Albert algorithm, the Pearson correlation coefficient between the mean path length and the required timesteps was -0.19, with a p-value less than 0.00001 which shows the correlation is statistically significant. The corresponding coefficient for Watts-Strogatz networks was -0.42 with a p-value less than 0.00001, which shows this correlation is also statistically significant.

\section{Discussion}
The model shows promising results in that its vulnerability is correlated with the connectedness of the network and the importance (eigenvector) of conspirators. 
Considering the minimum required time, the resistance of a population against conspiracies was imitated. The results show that a network that is built by the Watts-Strogatz algorithm was slightly more vulnerable to conspiracies than a network which is built by Barabasi-Albert algorithm (scale free). The reason for this difference might be that a Watts-Strogatz network has a high local clustering, while holding a short average path length like random networks, therefore the propagation of (dis)information is easier and faster in them. On the other hand, while noting that we used only 4 conspirators in our model, and the total number of agents were 100, the probability that conspirators become a hub was low, and therefore their ability to propagate their conspiracy was slightly smaller than in Watts-Strogatz network. A probable misinterpretation here might be that a less connected network is better against conspiracy theories. This is not correct because in our model, only conspirators were acting to deceive the population and were trying to propagate their disinformation in the whole network. In fact, the results of this study can be interpreted in a content-agnostic manner. In reality, both sides of the discussion try to influence the network. Nevertheless, both networks show that the connectedness and the eigenvector centrality of conspirators is a network is highly correlated with the network's vulnerability to conspiracies.

This was a first step to make a framework for computational study of conspiracy propagation. Further research must be conducted using other network formation algorithms. Other ranges of conspirators' ratio and population sizes should also be experimented. Another extension to this research might be to test the  effect  of  multi-dimensional  opinion-space in agents. Finally, other non-Bayesian models  of  opinion  dynamics could be  tested.

%
%
%

\end{document}